\documentclass{article}

    \PassOptionsToPackage{numbers, compress}{natbib}

\usepackage[preprint]{neurips_2019}

\usepackage[utf8]{inputenc} 
\usepackage[T1]{fontenc}    
\usepackage{hyperref}       
\usepackage{url}            
\usepackage{booktabs}       
\usepackage{amsfonts}       
\usepackage{nicefrac}       
\usepackage{microtype}      
\usepackage{multirow}
\usepackage{graphicx}
\usepackage{tipa}
\usepackage{todonotes} 
\usepackage{amsmath}

\newcommand{\appropto}{\mathrel{\vcenter{
  \offinterlineskip\halign{\hfil$##$\cr
    \propto\cr\noalign{\kern2pt}\sim\cr\noalign{\kern-2pt}}}}}

\title{A Surprising Density of Illusionable Natural Speech}

\author{%
  Melody Y. Guan \\
  Department of Computer Science\\
  Stanford University\\
  Stanford, CA 94305 \\
  \texttt{mguan@stanford.edu} \\
 \And
    Gregory Valiant \\
  Department of Computer Science\\
  Stanford University\\
  Stanford, CA 94305 \\
  \texttt{gvaliant@stanford.edu} \\
}

\begin{document}

\maketitle

\begin{abstract}
Recent work on adversarial examples has demonstrated that most natural inputs can be perturbed to fool even state-of-the-art machine learning systems. But does this happen for humans as well? In this work, we investigate \emph{what fraction of natural instances of speech can be turned into ``illusions'' which either alter humans' perception or result in different people having significantly different perceptions?} We first consider the McGurk effect, the phenomenon by which adding a carefully chosen video clip to the audio channel affects the viewer's perception of what is said \citep{mcgurk1976hearing}. We obtain empirical estimates that a significant fraction of both words and sentences occurring in natural speech have some susceptibility to this effect. We also learn models for predicting McGurk illusionability. Finally we demonstrate that the Yanny or Laurel auditory illusion \citep{pressnitzer2018auditory} is not an isolated occurrence by generating several very different new instances. We believe that the surprising density of illusionable natural speech warrants further investigation, from the perspectives of both security and cognitive science.
\end{abstract}

\section{Introduction}
\label{sec:Introduction}

A growing body of work on adversarial examples has identified that for machine-learning (ML) systems that operate on high-dimensional data, for nearly every natural input 
there exists a small perturbation of the point that will be misclassified by the system, posing a threat to its deployment in certain critical settings~\citep{behzadan2017vulnerability,chen2017ead,evtimov2017robust,huang2017adversarial,kurakin2016adversarial,papernot2017practical,papernot2016towards}. 
More broadly, the susceptibility of ML systems to adversarial examples has prompted a re-examination of whether current ML systems are truly learning or if they are assemblages of tricks that are effective yet brittle and easily fooled~\citep{samhita2013clever}. Implicit in this line of reasoning is the assumption that instances of ''real" learning, such as human cognition, yield extremely robust systems.  Indeed, at least in computer vision, human perception is regarded as the gold-standard for robustness to adversarial examples.


Evidently, humans can be fooled by a variety of \emph{illusions}, whether they be optical, auditory, or other; and there is a long line of research from the cognitive science and psychology communities investigating these \citep{hillis2002combining}.  In general, however, these illusions are viewed as isolated examples that do not arise frequently, and which are far from the instances encountered in everyday life.

\begin{figure}[!htbp]
\centering
\includegraphics[width=0.5\textwidth]{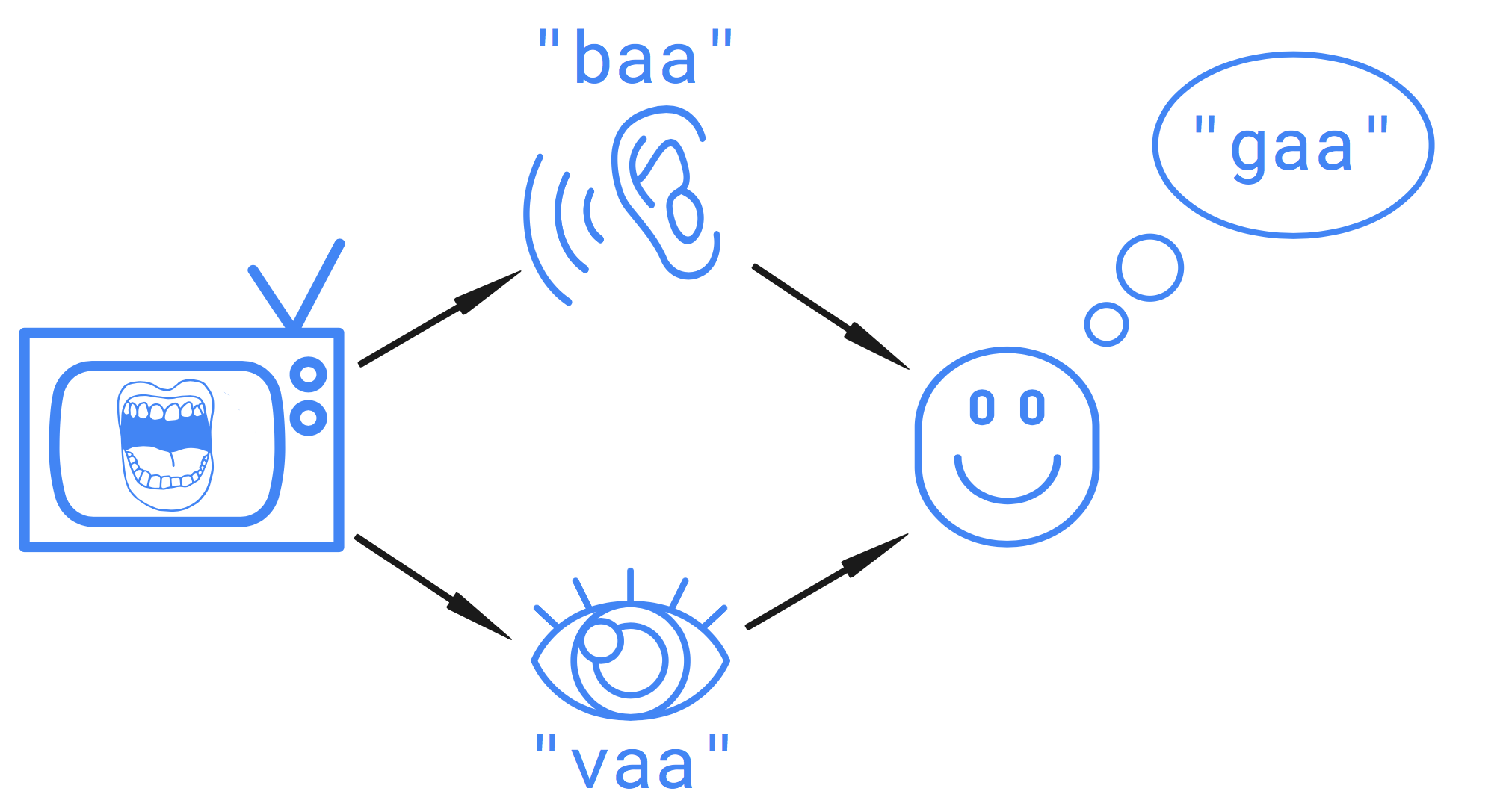}
\caption{Illustration of the McGurk effect. For some phoneme pairs, when the speaker visibly mouths phoneme $A$ but the auditory stimulus is actually the phoneme $B$, listeners tend to perceive a phoneme $C\neq B$.}
\label{fig:mcgurk}
\end{figure}

In this work, we attempt to understand how susceptible humans' perceptual systems for natural speech are to carefully designed ``adversarial attacks.'' We investigate the \textit{density} of certain classes of illusion, that is, the fraction of natural language utterances whose comprehension can be affected by the illusion.  Our study centers around the McGurk effect, which is the well-studied phenomenon by which the perception of what we hear can be influenced by what we see~\citep{mcgurk1976hearing}. 
A prototypical example is that the audio of the phoneme ``baa,'' accompanied by a video of someone mouthing ``vaa'', can be perceived as ``vaa'' or ``gaa'' (Figure \ref{fig:mcgurk}).  This effect persists even when the subject is aware of the setup, though the strength of the effect varies significantly across people and languages and with factors such as age, gender, and disorders~\citep{bastien2010audio,delbeuck2007alzheimer,irwin2006sex,mongillo2008audiovisual,norrix2006auditory,norrix2007auditory,sekiyama1997cultural,sekiyama2014enhanced,youse2004auditory}.

A significant density of illusionable instances for humans might present similar types of security risks as adversarial examples do for ML systems. Auditory signals such as public service announcements, instructions sent to first responders, etc., could be targeted by a malicious agent.  Given only access to a screen within eyesight of the intended victims, the agent might be able to significantly obfuscate or alter the message perceived by those who see the screen (even peripherally). 


\subsection{Related work}
\label{sec:related}

Illusionable instances for humans are similar to adversarial examples for ML systems. Strictly speaking, however, our investigation of the density of natural language for which McGurk illusions can be created, is not the human analog of adversarial examples. The adversarial examples for ML systems are datapoints that are misclassified, despite being extremely similar to a typical datapoint (that is correctly classified). Our illusions of misdubbed audio are not extremely close to any typically encountered input, since our McGurk samples have auditory signals corresponding to one phoneme/word and visual signals corresponding to another. Also, there is a compelling argument for why the McGurk confusion occurs, namely that human speech perception is bimodal (audio-visual) in nature when lip reading is available~\citep{besle2004bimodal,summerfield1992lipreading}.

To the best of our knowledge, prior to our work, there has been little systematic investigation of the extent to which the McGurk effect, or other types of illusions, can be made dense in the set of instances encountered in everyday life. The closest work is~\citet{elsayed2018adversarial}, where the authors demonstrate that some adversarial examples for computer vision systems also fool humans when humans were given less than a tenth of second to view the image. However, some of these examples seem less satisfying as the perturbation acts as a pixel-space interpolation between the original image and the ``incorrect'' class. This results in images that are visually borderline between two classes, and as such, do not provide a sense of illusion to the viewer. In general, researchers have not probed the robustness of human perception with the same tools, intent, or perspective, with which the security community is currently interrogating the robustness of ML systems.

\section{Problem setup}
\label{sec:Setup}

For the McGurk effect, we attempt an \textit{illusion} for a language token (e.g. phoneme, word, sentence) $x$ by creating a video where an audio stream of $x$ is visually dubbed over by a person saying $x'\neq x$.  We stress that the audio portion of the illusion is \emph{not} modified and corresponds to a person saying $x$.  The illusion  $f(x',x)$ \textit{affects} a listener if they perceive what is being said to be $y\neq x$ if they watched the illusory video whereas they perceive $x$ if they had either listened to the audio stream without watching the video or had watched the original unaltered video, depending on specification. We call a token \textit{illusionable} if an illusion can be made for the token that affects the perception of a significant fraction of people.

In Section \ref{sec:McGurk}, we analyze the extent to which the McGurk effect can be used to create illusions for phonemes, words, and sentences, and analyze the fraction of natural language that is susceptible to such illusionability. We thereby obtain a lower bound on the density of illusionable natural speech.

We find that 1) a significant fraction of words that occur in everyday speech can be turned into McGurk-style illusions, 2) such illusions persist when embedded within the context of natural sentences, and in fact affect a significant fraction of natural sentences, and 3) the illusionability of words and sentences can be predicted using features from natural language modeling. 

\section{McGurk experiments}
\label{sec:McGurk}

\subsection{Phoneme-level experiments} 
We began by determining which phoneme sounds can be paired with video dubs of other phonemes to effect a perceived phoneme that is different from the actual sound. We created McGurk videos for all vowel pairs preceded with the consonant /\textipa{n}/ as well as for all consonant pairs followed by the vowel /\textipa{a}/ spoken by a speaker. There are 20 vowel phonemes and 24 consonant phonemes in American English although /\textdyoghlig/ and /\textyogh/ are redundant for our purposes.\footnote{Refer to \url{www.macmillanenglish.com/pronunciation/phonemic-chart-in-american-english/} for a list of the  phoneme symbols of the International Phonetic Alphabet that are used in American English.} Based on labels provided by 10 individuals we found that although vowels were not easily confused, there are a number of illusionable consonants. We note that the illusionable phoneme pairs depend both on the speaker and listener identities. 

Given Table \ref{table:phonemes} of illusionable phonemes, the goal was then to understand whether these could be leveraged within words or sentences; and if so, the fraction of natural speech that is susceptible.  
\begin{table}[t]
\centering
\caption{Illusionable phonemes and effects based on preliminary phoneme-pair testing. Where a number of lip movements were available to affect a phoneme, the most effective one is listed.  
}
\label{table:phonemes}
\begin{tabular}{cccc}
    \hline
    \bf Phoneme & \bf Lip Movement & \bf Perceived Sound
    \\
    \hline
     \textipa{b}/ &/\textipa{w}/	&/\textipa{v}/, /\textipa{f}/, /\textipa{p}/ 
     \\
      \textipa{D}/ &/\textipa{b}/	&/\textipa{b}/ 
      \\
      \textipa{f}/ &/\textipa{z}/	&/\textipa{T}/, /\textipa{t}/, /\textipa{b}/ 
      \\
      \textipa{m}/ &/\textipa{D}/	&/\textipa{n}\textipa{T}/, /n/, /ml/ 
        \\
        \textipa{p}/ &/\textipa{t}/	&/\textipa{t}/, /\textipa{k}/ 
        \\
        \textipa{v}/ &/\textipa{b}/	&/\textipa{b}/ 
        \\
        \textipa{d}/ &/\textipa{v}/ &/\textipa{v}/, /\textipa{t}/
        \\
        \textipa{l}/&/\textipa{v}/&/\textipa{v}/
        \\
        \textipa{T}/&/\textipa{v}/&/\textipa{d}/, /\textipa{k}/, /\textipa{t}/, /\textipa{f}/
        \\
        \textipa{w}/&/\textipa{l}/&/\textipa{l}/
        \\
    \hline
\end{tabular}
\end{table}



\begin{table}[!htbp]
\centering
\caption{The 200 unique words sampled from the Project Gutenberg novel corpus. The 147 of those for which an illusory video was created are listed on top.  Ordering is otherwise alphabetical.}
\label{table:200}
\resizebox{.944\textwidth}{!}{
\begin{tabular}{lllllllll}
    \hline
    \bf \multirow{21}{*}{\rotatebox{90}{Illusion Attempted}} & about& addressed& all& also& and
    & anyone& arms\\
    & away& bad& be& been& before& behind& besides\\
    & blind& bought& box&  brothers& but& by& call\\
   & called& calling& came   & child& close& coming& could\\
   & days& dead & did& die& direction& done& else\\
	& end & even& everything& far& features & fell& few\\
    & fighting& fly& for& formed& from& game& gathered\\
    & gave& general& generally& god& good& half& hands\\
    & happened& hath& have& him& himself& idea& information\\
    & july& large& let& letter& life& like& list\\
    & made& many& mass& may& me& meet& men\\
    & months& more& Mrs& my& myself& never& nothing\\
    & of& off& old& one& open& opinion& ordinary\\
    & other& outside& passion& perhaps& please& plenty& point\\
    & possessed& present& put& questions& roof& said& save\\
    & seized& shall& sharp& ship& should& slow& some\\
    & speech& still& successful& summer& terms& than& that\\
    & the& their& them& themselves& there& they& things\\
    & though& time& top& upon& used& very& waited\\
    & was& water& we& went& what& when& which\\
    & will& wisdom& with& working& world& would& wounded\\
    \hline
    \bf \multirow{8}{*} {\rotatebox{90}{No Attempt}} & a& act& air& an& any& are& as\\
    & at& change& city& country& eyes& go& going \\
    & hair& has& he& heart& her& higher& his \\
    & house& i& in& into& is& it& its \\
    & king& know& nature& new& no& not& now \\
    & on& or& our& out& rest& saw& see \\
    & seen& she& sorrow& strange& take& talking& to \\
    & turn& who& writing& your&&&\\
    \hline
\end{tabular}
}
\end{table}

\subsection{Word-level experiments}
\label{sec:wordexperiments}


We sampled 200 unique words (listed in Table \ref{table:200}) from the 10,000 most common words in the Project Gutenberg novels in proportion to their frequency in the corpus. The 10k words collectively have a prevalence of 80.6\% in the corpus. Of the 200 sampled words, 147 (73.5\%) contained phonemes that  our preliminary phoneme study suggested might be illusionable. For these 147 words, we paired audio clips spoken by the speaker with illusory video dubs of the speaker saying the words with appropriately switched out phonemes. We tested these videos on 20 naive test subjects who did not participate in the preliminary study. Each subject watched half of the words and listened without video to the other half of the words, and were given the instructions: "Write down what you hear. What you hear may or may not be sensical. Also write down if a clip sounds unclear to you. Not that a clip may sound nonsensical but clear." Subjects were allowed up to three plays of each clip.

We found that watching the illusory videos led to an average miscomprehension rate of 24.8\%, a relative 148\% increase from the baseline of listening to the audio alone (Table \ref{table:word_results}).  The illusory videos made people less confident about their correct answers, with an additional 5.1\% of words being heard correctly but unclearly, compared to 2.1\% for audio only.  For 17\% of the 200 words, the illusory videos increased the error rates by more than 30\% above the audio-only baseline.

\begin{table}[t]
\centering
\caption{Test results for word-level McGurk illusions among the 147 words predicted to be illusionable.  Shown are average error rates for watching the illusory video vs listening to the audio only, as well as the percentage of words that are correctly identified but sound ambiguous to the listener.}
\label{table:word_results}
\begin{tabular}{cccc}
    \hline
    \bf  & \bf Audio Only & \bf Illusory Video & \bf Relative Increase\\
    \hline
    Words Incorrectly Identified 
    &10.0
    &24.8
    &	+148\% \\
            \hline
    Words Correct with Low Confidence 
    &2.1
    &5.1
    &	+144\% \\
        \hline
\end{tabular}
\end{table}

\subsubsection{Prediction model for word-level illusionability}

To create a predictive model for word-level illusionability, we used illusionable phonemes enriched with positional information as features. Explicitly, for each of the 10 illusionable phonemes, we created three features from the phoneme being in an initial position (being the first phoneme of the word), a medial position (having phonemes come before and after), or a final position (being the last phoneme of the word). We then represented each word with a binary bag-of-words model  \cite{joachims1998text}, giving each of the 30 phonemes-with-phonetic-context features a value of 1 if present in the word and 0 otherwise. We performed ridge regression on these features with a constant term. We searched for the optimal $l2$ regularization constant among the values [0.1, 1, 10, 100] and picked the optimal one based on training set performance. The train:test split was in the proportion 85\%:15\% and was randomly chosen for each trial. Across 10k randomized trials, we obtain average training and test set correlations of $91.1\pm 0.6\%$ and $44.6\pm 28.9\%$ respectively.

Our final model achieves an out-of-sample correlation of 57\% between predicted and observed illusionabilites. Here, the observed illusionability of the words is calculated as the difference between the accuracy of watchers of the illusory videos and the accuracy of listeners, where ``accuracy'' is defined as the fraction of respondents who were correct. For each word, the predicted illusionability is calculated from doing inference on that word using the averaged regression coefficients of the regression trials where the word is not in the training set.
 
\begin{figure}[!htbp]
\centering
\includegraphics[width=0.6\textwidth]{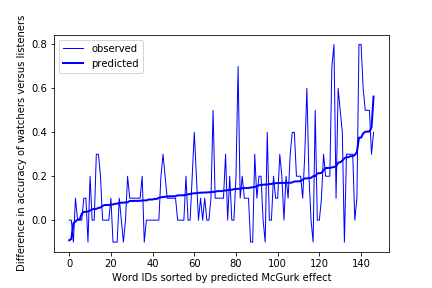}
\caption{Predicted word illusionability closely matches observed word illusionability, with out-of-sample correlation of 57\%. The words are sorted by increasing predicted word illusionability (and the observed illusionability of each word was \emph{not} used in calculating the prediction of that word). 
}
\label{fig:prediction}
\end{figure}

Our predicted illusionability is also \emph{calibrated}, in the sense that for the words predicted to have an illusionability <0.1, the mean empirical illusionability is 0.04; for words with predicted illusionability in the interval [0.1, 0.2] the mean empirical illusionability is 0.14; for predicted illusionability between [0.2, 0.3] the mean observed is  0.27; and for predicted illusionability >0.3, the mean observed is 0.50. Figure \ref{fig:prediction} visually depicts the match between the observed and predicted word illusionabilities.  

\subsection{Sentence-level experiments} 
\label{sec:sentenceexperiments}

We set up the following experiment on naturally occurring sentences. We randomly sampled 300 sentences of lengths 4-8 words inclusive from the novel \textit{Little Women}~\citep{alcott1994little} from the Project Gutenberg corpus. From this reduced sample, we selected and perturbed 32 sentences that we expected to be illusionable (listed in Table \ref{table:23}). With the speaker, we prepared two formats of each sentence: original video (with original audio), and illusory video (with original audio).
We then evaluated the perception of these on 1306 naive test subjects on Amazon Mechanical Turk.

The Turkers were shown videos for six randomly selected sentences, three illusory and three original, and were given the prompt: "Press any key to begin video [index \#] of 6. Watch the whole video, and then you will be prompted to write down what the speaker said." 
Only one viewing of any clip was allowed, to simulate the natural setting of observing a live audio/video stream. Each Turker was limited to six videos to reduce respondent fatigue. Turkers were also asked to report their level of confidence in what they heard on a scale from no uncertainty (0\%) to complete uncertainty (100\%). One hundred and twelve Turkers (8.6\%) did not adhere to the prompt, writing unrelated responses, and their results were omitted from analysis. 

We found that watching the illusory videos led to an average miscomprehension rate of 32.8\%, a relative 145\% increase from the baseline of watching the original videos (Table \ref{table:sentence_results}).  The illusory videos made people less confident about their correct answers. Turkers who correctly identified the audio message in an illusory video self-reported an average uncertainty of 42.9\%, which is a relative 123\% higher than the average reported by the Turkers who correctly understood the original videos. Examples of mistakes made by listeners of the illusory videos are shown in Table \ref{table:sentence_mistakes}. Overall we found that for 11.5\% of the 200 sampled sentences (23 out of the 30 videos we created), the illusory videos increased the error rates by more than 10\%. 

\begin{table}[!htbp]
\centering
\caption{The sentences randomly sampled from \emph{Little Women} for which we made illusory videos. Those that had observed illusionability >0.1 are listed on top.}
\label{table:23}
\begin{tabular}{lll}
    \hline
    \bf \multirow{12}{*}{Illusionable}
    & I'm not too proud to beg for Father.
    & No need of that.\\
    & I'd like to wear them, Mother can I?
    & It's no harm, Jo\\
   & Now do be still, and stop bothering.
   & Well, I like that!\\
	& How many did you have out?
	& Open the door, Amy!\\
    & Nonsense, that's of no use.
    & I am glad of that!\\
    & Of course that settled it.& I can't bear saints.\\
    & Serves me right for trying to be fine.& Of course I am!\\
    & There's gratitude for you!& That's my good girl.\\
    & Capital boys, aren't they?& You've got me, anyhow.\\
    & Brown, that is, sometimes.& I'll tell you some day.\\
    & What do you know about him?& He won't do that.\\
    & The plan of going over was not forgotten.\\
    \hline
    \bf \multirow{5}{*} {Not Illusionable}
    & On the way get these things.& I'm glad of it. \\
    & Aren't you going with him?& Then I'll tell you. \\
    & That was the question. & That's why I do it. \\
    & I hate to have odd gloves!& We don't know him! \\
    & What good times they had, to be sure\\
    \hline
\end{tabular}
\end{table}

\begin{table}[!htbp]
\centering
\caption{Test results for sentence-level McGurk illusions, showing average error rates for watching the illusory video vs watching the original video, and average self-reported uncertainty for correctly identified words.}
\label{table:sentence_results}
\begin{tabular}{cccc}
    \hline
    &\bf Original Video & \bf Illusory Video & \bf Relative Increase\\
    \hline
  Sentences Incorrectly Identified &13.4\%	&32.8\% &	+145\% \\
  Uncertainty of Correct Sentences &19.4\%	&42.9\% &	+121\% \\
        \hline
\end{tabular}
\end{table}

\begin{table}[!htbp]
\centering
\caption{Sample verbatim mistakes for sentence-level McGurk illusions. Spelling mistakes were ignored for accuracy calculations. Sample illusory videos are provided in the supplementary files.
}
\label{table:sentence_mistakes}
\begin{tabular}{ll}
    \hline
    \bf Sentence &  \bf Sample Listener Perceptions\\
    \hline
         Serves me right for trying to be fine.&Serves thee right for trying to be thine\\
      &  Serve you right for trying to be kind\\
    & serbes ye right shine by rhine\\
    \hline
    Of course that settled it. & up course bat saddle it\\
    & Of course Max settled it.\\
    \hline
     I can't bear saints. &I can't wear skates\\
    & I can't hear saints.\\
    & I can't spare sink\\
    \hline
     How many did you have out? &    How many did you knock out?\\
     & How many did you help out\\
    \hline
     I'll tell you some day. &I'll tell you Sunday.\\
     &  Ill tell you something\\
    \hline
    I'm not too proud to beg for Father. &I'm not too proud to beg or bother\\
    &I want the crowd to beg for father\\
    &Ine not too proud to beg for father\\
    \hline
        Well, I like that! &Well I like pets\\
    & Well, I like baths\\
    \hline
     Now do be still, and stop bothering. & now do we still end stop bothering\\
    & And do we still end sauce watering?\\
    \hline
     Of course I am! &Of course I an\\
     &  Of course I can\\
          &  Of course I anth.\\
    \hline
     You've got me, anyhow. &you got knee anyhow\\
     &  You got the anyhow.\\
    \hline
    There's gratitude for you! &Bears gratitude for you.\\
     &  Pairs gratitudes or you.\\
     & Bears gratitude lore you\\
    \hline
\end{tabular}
\end{table}

\subsubsection{Comparing word-level and sentence-level illusionabilities}
\label{sec:Comparison}

We obtained a sentence-level illusionability prediction model with an out-of-sample correlation of 33\% between predicted and observed illusionabilities. Here, the observed illusionability of the sentences was calculated as the difference between the accuracy of watchers of the illusory videos and the accuracy of watchers of original videos, where ``accuracy'' is defined as the fraction of respondents who were correct. We obtained predicted illusionabilities by simply using the maximum word illusionability prediction amongst the words in each sentence, with word predictions obtained from the word-level model. We attempted to improve our sentence-level predictive model by incorporating how likely the words appear under a natural language distribution, considering three classes of words: words in the sentence for which no illusion was attempted, the words for which an illusion was attempted, and the potentially perceived words for words for which an illusion was attempted. We used log word frequencies obtained from the the top 36.7k most common words from the Project Gutenberg corpus.\footnote{\url{https://en.wiktionary.org/wiki/Wiktionary:Frequency_lists\#Project_Gutenberg}} This approach could not attain better out-of-sample correlations than the naive method. This implies that context is important for sentence-level illusionability, and more complex language models should be used.

Finally, comparing word-level and sentence-level McGurk illusionabilities in natural speech, we observe that that the former is significantly higher. A greater McGurk effect at the word level is to be expected--sentences provide context with which the viewer could fill in confusions and misunderstandings. Furthermore, when watching a sentence video compared to a short word video, the viewer's attention is more likely to stray, both from the visual component of the video, which evidently reduces the McGurk effect, as well as the from the audio component, which likely prompts the viewer to rely even more heavily on context. Nevertheless, there remains a significant amount of illusionability at the sentence-level.

\section{Future Directions} 


This work is an initial step towards exploring the density of illusionable phenomena for humans. There are many natural directions for future work.  In the vein of further understanding McGurk-style illusions, it seems worth building more accurate predictive models for sentence-level effects, and further investigating the security risks posed by McGurk illusions.  For example, one concrete next step in understanding McGurk-style illusions would be to actually implement a system which takes an audio input, and outputs a video dub resulting in significant misunderstanding.  Such a system would need to combine a high-quality speech-to-video-synthesis system~\citep{suwajanakorn2017synthesizing,zakharov2019few}, with a fleshed-out language model and McGurk prediction model.

There is also the question of how to guard against ``attacks'' on human perception. For example, in the case of the McGurk effect, how can one rephrase a passage of text in such a way that the meaning is unchanged, but the rephrased text is significantly more robust to McGurk style manipulations? The central question in this direction is what fraction of natural language can be made robust without significantly changing the semantics.

A better understanding of when and why certain human perception systems are nonrobust can also be applied to make ML systems more robust. In particular, neural networks have been found to be susceptible to adversarial examples in automatic speech recognition \citep{schonherr2018adversarial,carlini2018audio} and to the McGurk effect~\citep{ngiam2011multimodal}, and a rudimentary approach to making language robust to the latter problem would be to use a reduced vocabulary that avoids words that score highly in our word-level illusionability prediction model. Relatedly, at the interface of cognitive science and adversarial examples, there has been work suggesting that humans can anticipate when or how machines will misclassify, including for adversarial examples \citep{chandrasekaran2017takes,Harding2017human,zhou2019humans}.


More broadly, as the tools for probing the weaknesses of ML systems develop further, 
it seems like a natural time to reexamine the supposed robustness of human perception. We anticipate unexpected findings. To provide one example, we  summarize some preliminary results on audio-only illusions.

\subsection{Auditory Illusions}
\label{sec:Future}

An audio clip of the word ``Laurel" gained widespread attention in 2018, with coverage by notable news outlets such as \textit{The New York Times} and \textit{Time}.  Roughly half of listeners perceive ``Laurel'' and the other half perceive ``Yanny'' or similar-sounding words, with high confidence on both sides \citep{pressnitzer2018auditory}. One of the reasons the public was intrigued is because examples of such phenomena are viewed as rare, isolated instances. 
In a preliminary attempt to investigate the density of such phenomena, we identified five additional distinct examples (Table \ref{table:yanny}). The supplementary files include 10 versions of one of these examples, where listeners tend to perceive either ``worlds'' or ``yikes.'' Across the different audio clips, one should be able to hear both interpretations. The threshold for switching from one interpretation to another differs from person to person.






\begin{table}[!htbp]
\centering
\caption{Test results for Yanny/Laurel style illusions.  Each row displays a cluster of similar-sounding reported words, with counts.  Emdashes indicate unintelligible.}
\label{table:yanny}
\resizebox{.55\textwidth}{!}{\begin{tabular}{lcll}
    \hline
    \bf Word & \bf Slowdown & \bf Perceived Sound & \bf N\\
    \hline
     worlds & 1.5x & worlds & 5 \\
     & & yikes/yites & 4 \\
     & &  nights/lights & 6 \\
    \hline
    bologna & 1.7x & bologna & 2 \\
     & & alarming & 2 \\
     & & alarmy/ayarmy/ignore me & 3 \\
     & & uoomi/ayomi/wyoming & 3 \\
     & & anomi/anolli/amomi & 3 \\
     & & -- & 2 \\
     \hline
     growing & 1.3x & growing & 8 \\
     & & pearling & 3 \\
     & & curling & 3 \\
     & & crowing & 1 \\
    \hline
     potent & 1.7x & potent/poatin/poden & 4 \\    
     & & pogie/bowie/po-ee & 3 \\
     & & pone/paam/paan & 3 \\
     & &  power/poder/pair & 3 \\
     & & tana & 1\\
     & & -- & 2 \\
    \hline
     prologue & 1.9x & prologue & 2 \\
     & & prelude/pro-why/pinelog & 3 \\
     & & kayak/kayank/kayan & 6 \\
     & & turnip/tienap/tarzan & 3 \\
     & & -- & 1 \\
    \hline
\end{tabular}}
\end{table}

These examples were generated by examining 5000 words, and selecting the 50 whose spectrograms contain a balance of high and low frequency components that most closely matched those for the word ``Laurel". Each audio file corresponded to the Google Cloud Text-to-Speech API synthesis of a word, after low frequencies were damped and the audio was slowed $1.3$-$1.9$x.  After listening to these top 50 candidates, we evaluated the most promising five on a set of 15 individuals (3 female, 12 male, age range 22-33).  We found multiple distributional modes of perceptions for all five audio clips. For example, a clip of ``worlds'' with the high frequencies damped and slowed down 1.5x was perceived by five listeners as ``worlds'', four as ``yikes/yites'' and six as ``nights/lights''. 
While these experiments do not demonstrate a density of such examples with respect to the set of all words---and it is unlikely that illusory audio tracks in this style can be created for the majority of words---they illustrate that even the surprising Yanny or Laurel phenomenon is not an isolated occurrence. It remains to be seen how dense such phenomena can be, given the right sort of subtle audio manipulation. 
    `
\section{Conclusion} 
\label{sec:Conclusion}

Our work suggests that for a significant fraction of natural speech,  human perception can be altered by using subtle, learnable perturbations. This is an initial step towards exploring the density of illusionable phenomenon for humans, and examining the extent to which human perception may be vulnerable to security risks like those that adversarial examples present for ML systems. 

We hope our work inspires future investigations into the discovery, generation, and quantification of multimodal and unimodal audiovisual and auditory illusions for humans. There exist many open research questions on when and why humans are susceptible to various types of illusions, how to model the illusionability of natural language, and how natural language can be made more robust to illusory perturbations. Additionally, we hope such investigations inform our interpretations of the strengths and weaknesses of current ML systems. Finally, there is the possibility that some vulnerability to carefully crafted adversarial examples may be inherent to all complex learning systems that interact with high-dimensional inputs in an environment with limited data; any thorough investigation of this question must also probe the human cognitive system.

\section{Acknowledgements}
This research was supported in part by NSF award AF:1813049, an ONR Young Investigator award (N00014-18-1-2295), and a grant from Stanford's Institute for Human-Centered AI. The authors would like to thank Jean Betterton, Shivam Garg, Noah Goodman, Kelvin Guu, Michelle Lee, Percy Liang, Aleksander Makelov, Jacob Plachta, Jacob Steinhardt, Jim Terry, and Alexander Whatley for useful feedback on the work. The research was done under Stanford IRB Protocol 46430.

\small
\bibliography{neurips_2019}
\bibliographystyle{plainnat}

\clearpage
\appendix

\end{document}